\documentclass[sigconf]{acmart}

\usepackage{xcolor,xspace,multirow,hyperref}
\AtBeginDocument{%
  \providecommand\BibTeX{{%
    \normalfont B\kern-0.5em{\scshape i\kern-0.25em b}\kern-0.8em\TeX}}}

\newcommand{\msec}[1]{\S\,\ref{#1}}

\newcommand{\auto}[1]{AutoML}
\newcommand{\revise}[1]{{#1}}

\newcommand{\add}[1]{{#1}}
\newcommand{\del}[1]{\iffalse{\textcolor{blue}{\sout{#1}}}\fi}
\newcommand{\rev}[2]{\iffalse{\textcolor{blue}{\sout{#1}}}\fi{#2}}

\copyrightyear{2023} 
\acmYear{2023} 
\setcopyright{acmlicensed}\acmConference[CHI '23]{Proceedings of the 2023 CHI Conference on Human Factors in Computing Systems}{April 23--28, 2023}{Hamburg, Germany}
\acmBooktitle{Proceedings of the 2023 CHI Conference on Human Factors in Computing Systems (CHI '23), April 23--28, 2023, Hamburg, Germany}
\acmPrice{15.00}
\acmDOI{10.1145/3544548.3581082}
\acmISBN{978-1-4503-9421-5/23/04}
\begin{document}

\title{AutoML in The Wild: Obstacles, Workarounds, and Expectations}


\author {Yuan Sun}
\email{yws5055@psu.edu}
\orcid{0000-0002-0752-1402}
\affiliation{%
  \institution{Pennsylvania State University}
  \city{University Park}
  \country{USA}
  }

\author {Qiurong Song}
   \email{qzs5098@psu.edu}
\orcid{0000-0001-9223-9593}
\affiliation{%
    \institution{Pennsylvania State University}
  \city{University Park}
  \country{USA}
}
  
 \author {Xininig Gui}
   \email{xinninggui@psu.edu}
   \orcid{0000-0002-9436-7940}
 \affiliation{%
  \institution{Pennsylvania State University}
    \city{University Park}
  \country{USA}
}
  
   \author {Fenglong Ma}
   \email{fenglong@psu.edu}
   \orcid{0000-0002-4999-0303}
\affiliation{%
  \institution{Pennsylvania State University}
    \city{University Park}
  \country{USA}
}
 
   \author {Ting Wang}
   \email{inbox.ting@gmail.com}
   \orcid{0000-0003-4927-5833}
\affiliation{%
  \institution{Pennsylvania State University}
    \city{University Park}
  \country{USA}
}

\begin{abstract}

Automated machine learning (AutoML) is envisioned to make ML techniques accessible to ordinary users. Recent work has investigated the role of humans in enhancing AutoML functionality throughout a standard ML workflow. However, it is also critical to understand how users adopt existing AutoML solutions in complex, real-world settings from a holistic perspective. To fill this gap, this study conducted semi-structured interviews of AutoML users ($N$ = 19) focusing on understanding (1) the limitations of AutoML encountered by users in their real-world practices, (2) the strategies users adopt to cope with such limitations, and (3) how the limitations and workarounds impact their use of AutoML. Our findings reveal that users actively exercise user agency to overcome three major challenges arising from customizability, transparency, and privacy. Furthermore, users make cautious decisions about whether and how to apply AutoML on a case-by-case basis. Finally, we derive design implications for developing future AutoML solutions.

\end{abstract}



\maketitle

\section{Introduction}
\begin{figure*}[!ht]
    \centering
    \includegraphics[width=0.9\textwidth]{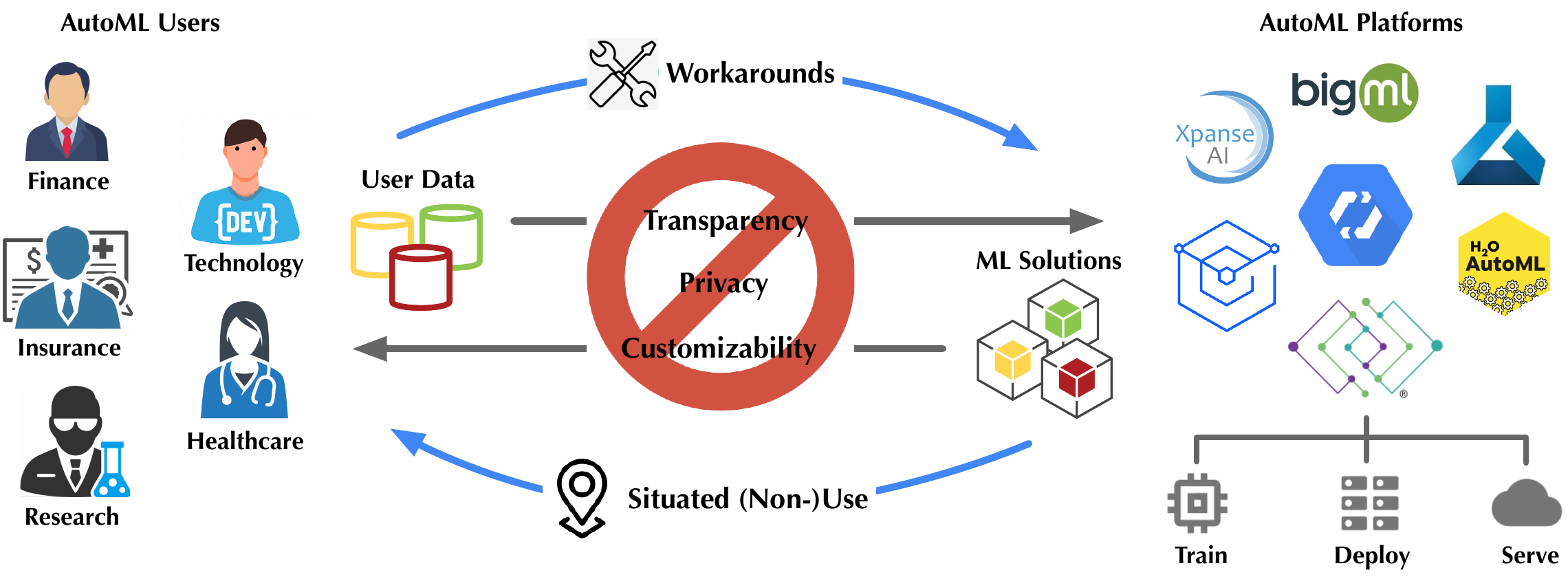}
    \caption{\add{Obstacles and Workarounds of Using AutoML in Complex, Real-world Settings.}}
    \label{fig:my_label}
\end{figure*}

While machine learning (ML) has been successfully applied to solve many challenging tasks across various domains, building performant ML solutions still requires substantial resources and extensive human expertise~\cite{he2021automl}. Automated machine learning (AutoML), a novel concept for automating the whole ML pipeline without (or as little as possible) human intervention~\cite{hutter2019automated}, has emerged as a way \add{to significantly reduce expensive development costs~\cite{tuggener2019automated}. As illustrated in Fig.~\ref{fig:my_label}, envisioned to enable domain experts without considerable ML backgrounds (e.g., marketing and business analysts)} to build ML solutions more easily, AutoML holds the promise of making ML techniques accessible to more people. Meanwhile, by liberating users from repetitive ML tasks (e.g., data preprocessing, parameter tuning, and feature selection), domain experts can spend more time on essential tasks, while data scientists can build more ML models in less time, improve model quality and accuracy, and experiment with more new algorithms. 

Despite its tremendous potential, the current discourse around AutoML is a mixture of hope and frustration. On one hand, AutoML is believed to be the driving force of ``democratizing data science''~\cite{shang2019democratizing}. On the other hand, the realization of automating ML workflows faces severe challenges~\cite{xin2021whither}, resulting in limited adoption of AutoML by practitioners~\cite{blom2021automl}. To facilitate the design of AutoML and support user experience, recent HCI studies have started investigating how users perceive and use AutoML in practice. One line of research focused on user trust in AutoML~\cite{vereschak2021evaluate, drozdal2020trust} and studied how to incorporate visualization tools to improve AutoML's transparency~\cite{wang2019atmseer,narkar2021model}. Another line of research gathered qualitative data about how users apply AutoML within the standard ML workflow to understand AutoML's benefits and deficiencies from the users' perspective~\cite{xin2021whither,crisan2021fits, musigmann2022diagonstic}. These studies examined the roles of users as data scientists throughout a standard computational pipeline and emphasized the importance of bringing ``human-in-the-loop'' to strike a balance between human control and machine automation~\cite{amershi2019software, liao2020questioning}. The primary goal was to compensate for automation-induced functionality deficiencies through human intervention to improve AutoML performance~\cite{passi2018trust, xin2021whither}. However, in complex, real-world settings, the concrete data science tasks are defined and executed by users with varying roles, interests, and backgrounds~\cite{passi2018trust}; users may face a variety of issues beyond AutoML's functionality, and they often define their goals according to complex situations and use different resources to accomplish them~\cite{suchman1987plans}. Therefore, using the standard ML workflow as a scaffold with the goal of achieving AutoML effectiveness may limit our understanding of how users perceive and use AutoML in practical settings, while relying primarily on a technical perspective of data science may fail to account for the ``social nuances, affective relationships, or ethical, value-driven concerns" of AutoML users~\cite{aragon2016developing}. This calls for a critical shift from examining the role of users in the standard ML workflow to understanding how users leverage AutoML as resources in their problem-solving processes.

The present study aims to fill this research gap by focusing on how users evaluate the role of AutoML in different situations and how they exercise user agency in leveraging AutoML in practice by developing various workaround strategies when facing challenges. As users' understanding of AutoML may be shaped by many situated factors, it is critical to investigate how users adopt and use AutoML in a variety of heterogeneous, specific situations. We conducted semi-structured interviews with 19 real-world AutoML users from different domains, varying in industries, job roles, and ML expertise levels. Specifically, our study discovered three major challenges, namely {\em lack of customizability}, {\em lack of transparency}, and {\em privacy concerns}, \add{which impede users from effectively applying AutoML in complex, practical situations, as illustrated in Fig.~\ref{fig:my_label}}. We analyzed the tactics devised by users to tackle such challenges and fit AutoML to diverse personal and organizational objectives, \add{including various workaround strategies as well as selective and situated (non-)use of AutoML.} Theoretical and design implications are provided.

\section{RELATED WORK}

\subsection{\add{HCI Research on AutoML}}


\subsubsection{\add{Users' Perceptions of AutoML}}

Recent studies have examined users' overall perceptions of AutoML. Wang et al.~\cite{wang2019human} focused on data scientists' perceptions of AutoML's utility and potential impacts on data science practices and occupations. Through a qualitative interview, they found that most data scientists consider AutoML as a complementary tool and perceive that AutoML is not able to replace human expertise. In another study, Wang et al.~\cite{wang2021autods} developed an experimental AutoML system that only requires data scientists to upload their datasets and generates ML models automatically, and conducted a user study on data scientists' experience with such systems. They found that although the system provides high-quality models in less time, data scientists show less trust in the AutoML-generated models, largely due to a lack of transparency. 

\subsubsection{\add{AutoML's Transparency}}

Many studies have focused on enhancing AutoML's transparency to foster user trust. Drozdal et al.~\cite{drozdal2020trust} found that including transparency features such as model performance metrics and visualization is critical for increasing user trust. In a similar vein, Wang et al.~\cite{wang2019atmseer} designed and implemented a visualization tool to support user understanding of the generated ML models and search space, and demonstrated that such visualization tools enhance user trust and enable users to better apply AutoML. Weidele et al.~\cite{weidele2020autoaiviz} developed an experimental system to visualize AutoML's model generation process and found that opening this black box improves user trust. Narkar et al.~\cite{narkar2021model} designed a visualization tool to support user decision-making by analyzing AutoML's outcomes and comparing candidate ML models based on multiple performance metrics. In addition to this strand of design and experimental effort, Crisan and Fiore-Gartland~\cite{crisan2021fits} interviewed data scientists in enterprise settings, focusing on how they use visualization as a means to integrate humans into the automation loop. However, they found that participants' use of visualization is fairly limited due to the lack of usefulness and benefits~\cite{crisan2021fits}. 

\subsubsection{\add{Human-in-The-Loop AutoML}}
Although AutoML's goal is to ``reduce the role of humans in the loop and fill the gap for non-expert users by playing the role of the domain expert''~\cite{elshawi2019automated}, there is a growing line of research that emphasizes the importance of ``human-in-the-loop''.\del{Specifically, human-in-the-loop in ML refers to humans and ML processes interacting to solve a problem~\cite{monarch2021human, lee2020human}. Bringing human knowledge into the ML workflow contributes to the learning process in many ways. Ideally,} 
\add{Intuitively, AutoML with humans in the loop enables users to monitor and control AutoML's different stages yet without manually taking over the whole process~\cite{crisan2021fits}, which also allows users to incorporate intuition and domain knowledge into AutoML's workflow to enhance its performance~\cite{gil2019towards, lee2020human}.} To this end, Behnaz et al.~\cite{arzani2021interpretable} designed a new AutoML framework with interpretable feedback that allows users to leverage their domain knowledge.\del{Purwanto et al.~\cite{Purwanto2021ManVM} compared AutoML-generated models with manually crafted models and found that experts with domain knowledge are still needed in the loop.} Gil et al.~\cite{gil2019towards} proposed a hybrid framework that provides an intelligent interface enabling users to specify problem settings and explore different models, while AutoML functions under users' guidance. Besides, there have been studies focusing on understanding the roles of humans and automation in the standard ML workflow from a ``human-in-the-loop'' perspective.\del{Previous HCI research on data science has uncovered the data science workflow. For instance, Muller et al.~\cite{muller2019data} found that data science workflow can be divided into four stages, including data acquisition, data cleaning and integrating, feature engineering, and model building and selection. Wang et al.~\cite{wang2019human} further offered a more detailed description of a data science work, including preparation (data acquisition, data cleaning, and labeling, feature engineering), modeling (model selection, hyperparameter optimization, ensembling, model validation), and deployment (model deployment, runtime monitoring, model improvement).}\del{Building upon such work, some studies focus on workflow and task-driven analysis of the roles of humans in the loop and the roles of automation in ML workflows. For instance,} Wang et al.~\cite{wang2021much} found that different data scientists desire for varying levels of automation depending on their experiences and the workflow stages. Crisan and Fiore-Gartland~\cite{crisan2021fits} further detailed the need for human involvement in different stages such as data preparation, analysis, deployment, and communication~\cite{muller2019data,wang2019human}. In a similar vein, Xin et al.~\cite{xin2021whither} examined user-reported AutoML's benefits and deficiencies and highlighted the importance of including humans in AutoML's process to combat AutoML's deficiencies such as system failures, lack of customizability, and lack of transparency.


\subsection{\add{User Agency and AutoML}}

\del{The recent AutoML work, especially those focused on transparency and human-in-the-loop, reflects the effort in balancing human agency and automation. Humans have the conscious hold the ``free will'' to formulate acts to influence our environment~\cite{haggard2002voluntary}. From the human agency perspective, humans are relatively free to enact technologies in different ways. Therefore, humans are capable of creating a wide variety of social meanings and enabling a vast range of possible implementations~\cite{orlikowski1992duality}. Because users operate in particular settings and develop extensive practical experience engaging with these tools, they may adjust their usage of the toolkits in ways that were not imagined by designers~\cite{desanctis1994capturing}. User action is not limited by the deficiencies of technology, and human agents may be resourceful enough to overcome technology's material constraints, making technology adaptable~\cite{boudreau2005enacting}.} 

\subsubsection{\add{Human Agency and Automation}}
\add{The aforementioned studies, from a broader view, reflect the desiderata to balance human agency and machine automation.}\del{In HCI, user agency is a vital consideration when evaluating how people experience automation.} While machine automation outperforms humans in terms of efficiency and cost reduction, over-reliance on automation may sacrifice humans' critical engagement and domain expertise~\cite{heer2019agency}. The tension between human agency and automation creates critical challenges for system design~\cite{heer2019agency}. HCI research has long advocated for more human-centered approaches that balance humans and  automation~\cite{heer2019agency,shneiderman1997direct,amershi2019guidelines,shneiderman2004agency}. For instance, the debate over ``direct manipulation versus interface agents'' between HCI and AI researchers back in the 1990s arrived at the conclusion of ``increased automation that amplifies the productivity of users and gives them increased capabilities in carrying out their tasks while preserving their sense of control and their responsibility''~\cite{shneiderman1997direct}. Along this line, Horvitz~\cite{horvitz1999principles} proposed a set of principles for designing systems to support automation-human collaboration efficiently, such as ``considering uncertainty about a user's goals'' and ``providing mechanisms for efficient agent-user collaboration to refine results''. Recently, by synthesizing AI design from various sources, Amershi et al.~\cite{amershi2019guidelines} proposed a set of guidelines to support the interaction between humans and AI.

\subsubsection{\add{Supporting User Agency in ML}} In the context of ML, existing work has investigated how to support user agency through interface mechanisms such as allowing user feedback~\cite{holzinger2016towards}, \add{improving customizability for ML tools~\cite{winter2020flatpack}, and enabling users to interactively modify AutoML's search space~\cite{wang2019atmseer}. Through self-generated actions, these technological affordances have been shown to effectively enhance user agency~\cite{sundar2010personalization}.}\del{that invite user actions and enhance the sense of agency through self-generated actions.} However, it is found that users often exert their own agency to modify or adapt ML solutions. For instance, Cai et al~\cite{cai2019human} designed refinement tools to assist medical decision-making and found that users often invent new strategies such as disambiguating ML and human errors to better understand the underlying algorithms. In a similar vein, Xin et al.~\cite{xin2021whither} showed that users switch back to manually developed ML models when they perceive high risks of using AutoML due to its lack of transparency. These ``workarounds'' are conscious and creative acts to ensure ML works in practice~\cite{alter2014workarounds}. Recent research ~\cite{gillies2016human} has pointed out that there is a lack of understanding of the nature of applying ML as a co-adaptive process, in which users adapt to using ML more effectively and adjust their goals accordingly.

\subsection{\add{Summary}} \add{While recognizing the importance of the role of humans in AutoML, existing HCI studies mainly focused on how humans may compensate for AutoML's deficiencies in each step of the standard ML workflow, such as developing interface affordances of customization.} However, in practice,\del{Humans have the conscious hold the ``free will'' to formulate acts to influence our environment~\cite{haggard2002voluntary}. From the human agency perspective, humans are relatively free to enact technologies in different ways. Therefore, humans are capable of creating a wide variety of social meanings and enabling a vast range of possible implementations~\cite{orlikowski1992duality}.} because users operate in complex, real-world settings, \add{have diverse expertise and backgrounds,} and develop varying experiences engaging with AutoML, they may adjust their use of AutoML in ways unimaginable by designers~\cite{desanctis1994capturing}; user actions are not limited by AutoML's deficiencies, and users may be resourceful enough to work around such limitations, fitting AutoML to their targets and needs ~\cite{boudreau2005enacting}. Built upon the strand of existing research, \add{our work takes on a more ``user-centered'' perspective,} with a special focus on how \add{real-world} users leverage AutoML as one of the resources in their problem-solving processes and the social-technical implications of AutoML in their practices, which contributes to the understanding of how applying AutoML is also a co-adaptive process.

\section{Method}
We conducted semi-structured interviews to understand users' perceptions of AutoML and their exercises of human agency in working with and around AutoML. We further identified user expectations for the future design of AutoML.

\subsection{Recruitment and Interviews}
We focused on users who have hands-on experience with AutoML in different domains. We recruited participants by spreading recruitment messages through words of month ($n$ = 10), mailing lists within enterprises ($n$ = 4), and social media ($n$ = 5). Participants were invited to complete a screening questionnaire about whether they had used AutoML before, which AutoML platforms they had used, and how long their experience was. We recruited 19 participants who reported having experience with AutoML in their workplaces. 

The interviews were conducted remotely from May 2022 to July 2022 after receiving the institutional review board (IRB) approval. Each interview was scheduled for 60 minutes on video conferencing platforms and was audio-recorded for transcription purposes. \add{The average interview duration was 45 minutes, with individual ones varying from 35 to 60 minutes.} We recruited participants who span a diverse range of domains across healthcare \add{($n$ = 3)}, finance \add{($n$ = 1)}, human resources \add{($n$ = 1)}, technologies \add{($n$ = 9)}, and academic research \add{($n$ = 3)}. In addition, participants have varying job roles, from the marketing manager at a social media agency to the head of technology at a healthcare company. The majority of our participants are based in the United States, with one based in China (P13), and one in Kenya (P6). Each participant received a \$50 gift card upon completion of the interview. A summary of participants' ML expertise levels, industries, job roles, and the type(s) of AutoML platforms they used is presented in Table~\ref{tab:my_Table}. We omitted the specific names of AutoML platforms to preserve the anonymity of participants. 

\begin{table*}{\small
    \centering
    \renewcommand{\arraystretch}{1.2}
    \setlength{\tabcolsep}{3pt}
    \caption{Characteristics of Participants in Our Study}
    \label{tab:my_Table}
    \begin{tabular}{ccccccc}
    \hline
{\bf Participant} & {\bf Gender} & {\bf ML Exp} & {\bf ML Job Role} & {\bf Industry} & {\bf Organization Size} &  {\bf AutoML Platform} \\
\hline
    P1 & Female & 2 yrs & Marketing Manager & Marketing & 10 - 100 & Commercial  \\
    P2 & Female & 5 yrs& Software Engineer & Finance & 1,000 - 10,000 & Commercial \\ 
    P3 & Male & 8 yrs& ML Engineer & Social Network & 1,000 - 10,000 & Internal \& Commercial \\ 
    P4 & Male & 3 yrs& ML Researcher & Healthcare & 100 -1,000 &Internal \\ 
    P5 & Male & 10 yrs & NLP Researcher & Human Resource & 10 - 100 & Commercial \\ 
    P6 & Male & 12 yrs& Head of Technology & Healthcare & 100 - 1,000 & Commercial \\ 
    P7 & Female & 3 yrs& HCI Researcher & University & 1,000 - 10,000 & Commercial  \\
    P8 & Male & 2 yrs& ML Researcher & Social Network & 1,000 - 10,000 & Internal \\
    P9 & Male & 15 yrs& Software Engineer & Information System & 50,000+ & Internal \\
    P10 & Female & 6 yrs& Researcher & University & 1,000 - 10,000 & Internal \\
    P11 & Male & 3 yrs& Software Engineer & Mobility Service & 1,000 - 10,000 & Internal \\
    P12 & Male & 4 yrs& Software Engineer & Technology & 50,000+ & Internal \\ 
    P13 & Male & 16 yrs& Research Manager & Technology & 50,000+ & Internal \\
    P14 & Male & 2 yrs& Data Scientist & Travel Technology & 1,000 - 10,000 & Internal \\
    P15 & Male & 5 yrs& ML Researcher & Retail & 1,000 - 10,000 & Commercial\\
    P16 & Male & 2 yrs& Data Scientist & Healthcare & 1,000 - 10,000 & Commercial \\
    P17 & Male & 3 yrs& Researcher & University & 1,000 - 10,000 & Internal \\
    P18 & Male & 5 yrs& Software Engineer & Social Network & 1,000 - 10,000 & Commercial\\
    P19 & Male & 2 yrs& ML Researcher & Music Streaming Service & 1,000 - 10,000 & Internal \\\hline
    \end{tabular}}
\end{table*}

\subsection{Data Analysis}
The dataset for analysis included all 19 interview transcripts. \add{Two researchers (including the first author) manually transcribed the interviews. To ensure transcription accuracy, we carefully examined the data by repeatedly checking back against the original audio recordings. To provide contextual information, each interview began with open-ended questions: (i) can you tell us about the company/industry you are working in? (ii) can you tell us your current job responsibilities? (iii) how long have you been working in ML-related work? The contextual information helped the transcribers to interpret recordings if they were not the researchers who collected the data.} 

To discover the main themes of the interviews, we followed an inductive approach~\cite{thomas2003inductive} to perform thematic analysis~\cite{braun2006using, braun2019reflecting}. Four \add{trained} researchers \add{(two HCI and two ML researchers)} were involved in the data analysis. In the first stage, \add{by reading the transcripts independently and repeatedly, we actively searched for the meanings and patterns of the content and wrote down analytic memos. Through this iterative process, we became familiar with every aspect of the data without being selective or skipping over the data following the instructions of thematic analysis~\cite{braun2006using}.}

\add{After obtaining the initial understanding of how users apply AutoML in workplaces}, we conducted multiple rounds of discussions about our understanding based on the analytic memos. \add{Following a constant comparative approach~\cite{glaser2017discovery}, the process involved moving back and forth between the similarities and differences of emerging categories with reference to the collected data.} After that, we individually returned to the data and began assigning basic codes to each idea. \add{In this stage, each researcher coded the data by highlighting and noting the texts to indicate potential patterns. We followed the guidance of coding ``as many potential themes/patterns as possible''} ~\cite{braun2006using} and generated a list of 209 basic codes. \add{For example, the data extract ``we mainly look at how the three learning curves of training, validation, and testing change during the training process and the testing process'' was first identified as one of ``workarounds for lacking transparency'', and further coded as a sub-category of ``tracking AutoML's process'' of this category. \revise{We held regular meetings to discuss and compare the respective codes to note similarities and discrepancies. We held six meetings, each lasting an hour, to address any disagreements related to coding. We used the "Open Discussion" method ~\cite{chinh2019ways} to resolve these disagreements. During these meetings, we created a table that summarized the codes used by each of the four coders for each quote. The main goal was to discuss and resolve any discrepancies. The coders addressed each disagreement in the order that it appeared on the summary table. Before making a final decision, the coders considered the codes used by other coders for a particular quote and took into account each coder's rationale for using a specific code.} Note that we employed codes to facilitate the theory-development process, and avoided relying on inter-coder reliability to ensure all instances of variations can be captured and to prevent potential marginalization of viewpoints~\cite{mcdonald2019reliability}.}

\add{Upon generating the initial coding,} we reconvened to compare and discuss the codes and explain how each basic code can be used to represent a potential theme. \del{giving in a list that was unified with 209 basic codes.} We then analyzed the codes and decided how different codes can be combined to form a higher-level theme through multiple rounds of discussions. After that, we re-examined the candidate themes and refined the themes to ensure internal homogeneity and external heterogeneity~\cite{patton1990qualitative}. Lastly, we defined and named the themes and conducted multiple rounds of refinements before generating the final reports. \add{The final satisfactory thematic map includes two primary themes: ``acknowledging and working around AutoML's limitations'' and ``applying AutoML selectively and situationally''.}

\begin{table*}{\small 
    \centering
        \renewcommand{\arraystretch}{1.2}
    \caption{\add{Summary of Our Findings}} 
    \label{tab:findings}
    \begin{tabular}{cll}
    \hline
{\bf Challenge} & {\bf Workaround/Strategy} & {\bf Participant}  \\
\hline
\hline
\multirow{3}{*}{Customizability (\msec{sec:custom})} &  Contextualizing input data & P6, P7\\
& Incorporating domain knowledge & P3, P9, P13, P15 \\
& Building internal AutoML tools &  P6, P11, P17\\
    \hline
\multirow{3}{*}{Transparency (\msec{sec:transparency})} 
& Validating AutoML's outcomes manually & P2, P10, P16\\
& Tracking AutoML's process & P3, P8 \\ 
& Creating customized visualization & P1, P2, P3, P5, P13, P15\\
    \hline
\multirow{4}{*}{Privacy (\msec{sec:privacy})} 
&  Uprooting privacy leakage & P6, P10, P13, P15, P17, P18, P19\\
& Applying privacy-preserving techniques & P3, P9 \\
& Delegating to legal regulation & P1, P13\\ 
& Choosing trustworthy platforms & P2, P4, P5, P8, P9, P17\\
    \hline
    \hline
\multirow{3}{*}{Use vs. Non-Use (\msec{sec:selective})} & Performance-driven (non-)use & P1, P3, P5, P6, P7, P8\\
& Task-oriented (non-)use & P1, P2, P4, P7, P8, P10, P13, P17, P18\\
& Context-specific (non-)use & P3, P9, P14, P15, P17\\ 
    \hline
    \end{tabular}}
\end{table*}

\section{Findings}
Overall, our study found that participants are well aware of AutoML's inadequacies, including incompatibility with specific task contexts, lack of transparency, and potential privacy issues. However, rather than being impeded by its limitations, they set clear objectives for what AutoML is able to achieve and thus adapt their use accordingly. In practice, they adopt many strategies to cope with AutoML's inadequacies to maximize its practical usability. \add{Our findings are summarized in Table~\ref{tab:findings}.}

\subsection{Acknowledging and Working around AutoML's Limitations}

\subsubsection{\bf Tackling Lack of Customizability}
\label{sec:custom}
Requesting for more customizability is a sentiment shared by participants. \add{Both P7 and P9 pointed out that existing AutoML platforms are often encapsulated, making it difficult for users to intervene with the automation process or perform fine-grained tuning of the generated results.}

\begin{quote}
    \emph{``There is actually nothing, not so much you can do in, you know, adjusting the parameters, or whatever algorithm they use.''} (P7)
    
    \emph{``The cloud-based AutoML doesn't allow the user to export the model or download the model to deploy it on their own machines. There is no such feature now because they are using the most advanced models. Those models are corporate properties and should not be disclosed to anyone else.'' \del{It took millions of dollars to train these models. They don't want other companies to get it, and even if the user downloads them, they may not have the resources to deploy it because the same model may expand to multiple servers. They can only be run with clusters of servers just to make inferences.}} (P9)
\end{quote}

Moreover, we found that AutoML's lack of customizability is multi-fold and participants need to derive various strategies to tackle this challenge in different scenarios. \del{Below, we detail the multi-faced, lack of customizability issue, and the corresponding strategies that our participants adopted.} 

\paragraph{\underline{Workaround 1: \add{Contextualizing} Input Data}}
As most current AutoML platforms \add{are generic and do not provide the flexibility to configure their inner workings, participants often find lacking the capability to handle context-sensitive tasks.\del{To customize the AutoML system,} One common workaround is to contextualize the input data by adding ``context hints''}, so that AutoML is able to utilize such additional information to generate context-specific ML solutions (P6, P7). For instance, P7, who is an HCI researcher, conducted a user study to understand the user experience with a voice-based self-tracking application. Due to her limited coding experience, she chose a commercial AutoML platform to provide the natural language processing (NLP) functionality. However, she reflected that the platform was not adaptive to the self-tracking context, and she needed to add additional contextual information to the input data for AutoML to work precisely:
\begin{quote}
    \emph{``I feel that the AutoML services are not smart enough if you don't give them enough contextual information, they cannot accurately recognize users' voice input. I don't know how to improve that, so I tell my participants to give the system a little bit more contextual information. For example, ``7 to 9'' is often mistakenly translated into ``729'', and I asked my participants to say ``7 to 9 AM'' or ``7 to 9 in the morning'' so that it can help these systems improve their performance.''} (P7)
\end{quote}

\del{In P7's case, the AutoML platform that she used did not provide much flexibility in terms of tuning the models, but what data were taken as input could be partially controlled. When collecting data for model input, P7 specifically required the participants to provide more contextual information to help the system understand the context, in the hope of improving the AutoML service that can be better adapted to the context of self-tracking.}

\paragraph{\underline{Workaround 2: Incorporating Domain Knowledge}}
Another limitation related to AutoML's customizability perceived by participants is that it does not naturally fit the needs of different industries. Participants' workaround is to gather and incorporate domain knowledge into AutoML's optimization objectives (P3, P9, P13, P15). For example, P13, who works in a technology company focusing on providing AI and ML solutions to traditional industries, reported that AutoML was too generic and regarded it as \emph{``a product made by obtaining the greatest common divisor among the needs of all users.''} To fit AutoML to industry-specific tasks, he communicated with industry experts and transformed the experts' domain knowledge into AutoML's optimization objectives:

\begin{quote}
    \emph{``Based on our cooperation with enterprises in traditional industries, the most difficult but valuable thing is how to convert domain knowledge of different industries into your model design. It's actually the most valuable part, but this is definitely something I can't do with AutoML. For example, in the optimization of the supply chain, a relatively reasonable level of inventory should be maintained, if you have no one to tell you about this kind of domain knowledge, you can not make AutoML fit into this specific task. Since either ML or data scientists are not particularly familiar with such issues, it actually requires us to have more communication with industry experts and transform this kind of domain knowledge into the objective in my model, so the bridging work is actually very important.''} (P13)
\end{quote}

\del{P13 emphasized the importance of customizing the model design to fit the unique needs of specific industries as models should be domain-specific. However, AutoML cannot automatically deal with cases that require domain knowledge. Therefore, P13 customized AutoML by bridging the ML knowledge and industry-specific domain knowledge and defining AutoML tasks by incorporating the domain knowledge gathered from experts, so that the AutoML could better adapt to specific industries' needs. In this case, P13 played an important role in gathering and translating domain knowledge into model design.} 

\paragraph{\underline{Workaround 3: Building Internal AutoML Tools}}
AutoML's lack of customizability is also manifested in its limited support for uncommon data types. Correspondingly, participants often opt to build their own AutoML tools (P6, P11, P17). For instance, P11 explained that his company has developed AutoML tools to support tabular data, which is missing on mainstream platforms: 
\begin{quote}
   \emph{``Our company has developed our own AutoML platform. The AutoML platforms provided by companies like Google and Amazon are very mature. However, the functions of their AutoML platforms support generic data such as images and text but do not support tabular data, which our company deals with.''\del{That's why we develop our own platform to deal with tabular data.}} (P11) 
\end{quote}
\del{P11 pointed out their company's needs for processing tabular data, while the existing third-party AutoML tools may be better deployed for image and textual data but do not support tabular data well. Therefore, the company has developed its own AutoML tool that is customized to fit its own needs.}

\add{Similarly, P17 described that in his company the data comes in different formats and with different features requiring refining the search space, while current AutoML platforms do not provide such configurability:}  
\begin{quote}
   \emph{``Because the data in our field can be in many forms and has discrete features, it needs a better representation of the overall data. The process of its correction also needs to be searched. Our (internal) AutoML is designed to be more refined and can handle different kinds of input data.''} (P17)
\end{quote}

As another example, P6 works at a non-governmental organization (NGO) in Kenya. As the company provides healthcare information and helps patients connect with local medical resources, the ML solution needs to support \add{the local language} of Swahili. Therefore, the company is building an internal AutoML platform and is going to switch from the commercial AutoML service to its own platform, which can better support Swahili without sacrificing accuracy due to translation, as well as ``{\em significantly saves money for the company}.'' 

In summary, building internal AutoML tools is the strategy to deal with special data types or data with unique features, or to provide better-localized solutions.\del{to improve the compatibility of AutoML techniques in applied data science.}  

\subsubsection{\bf Tackling Lack of Transparency}
\label{sec:transparency}
The lack of transparency is another major concern frequently mentioned by participants. For example, P13 and P17 emphasized that while ML is already a black-box, automating ML adds another layer of ``black-boxness''; thus, they perceived AutoML as a ``double black-box''.\del{``{\em ML has created the blac-kbox problem,} {\em and AutoML just creates another black box out of it, just like a double black-box}.'' (P13)} The main transparency issues perceived by participants include two aspects: (i) AutoML has limited support to evaluate its \emph{outcomes}; and (ii) it also falls short to provide sufficient information to assess its \emph{process}. Thus, participants have devised various workarounds to assess and evaluate AutoML's outcomes and process.

\paragraph{\underline{Workaround 1: \rev{Evaluating the results}{Validating AutoML's Outcomes} Manually}} 
Several participants (P2, P10, P16) shared their struggles with evaluating and validating AutoML's \emph{outcomes}. For example, P16 pointed out the lack of indicative performance metrics on the AutoML platform he has used:

\begin{quote}
    \emph{``There is one issue with AutoML, at least according to our experience when cooperating with the NGO. The evaluation metrics it (AutoML) gave were relatively limited. I remember that it only had one metric of `precision' for classification, but other metrics such as `F1 score' and `accuracy' were missing.''} (P16)
\end{quote}
\del{According to P16, certain performance metrics were missing for them to evaluate the results, and the AutoML tool only provided the metric of precision, which was not sufficient to support the evaluation. F1 score is one of the metrics for evaluating models requiring a balance between precision and recall, and there's an uneven class distribution.} 

To cope with such transparency issues,\del{like this, our participants came up with various ways of evaluating the results.} one common workaround by participants is to manually validate AutoML's outcomes using self-selected metrics or checking their backward compatibility with existing ML solutions:\del{As the provided metrics of AutoML tools were usually limited, when our participants needed different performance metrics for different tasks, they would check them manually. For example, P10 shared,}
\begin{quote}
    \emph{``If it's just for the classification, I would just use the provided evaluation metrics like accuracy and some kind of like F1 score precision and recall, some kind of provided metrics; but for tasks like regression, I usually manually check whether the results are reliable.''} (P10)
    
    \emph{``In our company, we compare AutoML's results with the previous results. For example, when we want to run a credit score, we first use a well-trained model like our previous model to run to get a batch of results. Then we use AutoML to run the score. How much we can trust AutoML results depends on how different its results are from our previous results. If there is a big difference, there may be problems. If AutoML's results are within our acceptable range, there should be no big problem, but we definitely do a lot of such testing.''} (P2)
\end{quote}

\del{P2 pointed out that in their work, they compared the results of AutoML with the results of a previous well-trained model. Through such testing and comparison, they could evaluate the performance of the AutoML models based on their previous work experience and previous models to see whether AutoML results were trustworthy.}

\paragraph{\underline{Workaround 2: Tracking AutoML's Process}}
Further, most current platforms only provide explanations for AutoML's outcomes (e.g., the importance of different features for the models suggested by AutoML), while its dynamic process (e.g., how the models are actually found) remains fairly vague. However, as several participants (P3, P8) indicated, they equally care about evaluating the dynamics of AutoML's \emph{process} to assess whether it performs as expected. 

The reasons for this lack of process transparency on existing AutoML platforms may be multi-fold. For instance, commercial platforms often view the underlying AutoML techniques as proprietary intellectual properties and are unwilling to disclose the internal information. Also, as it often requires sufficient expertise to apprehend AutoML's process, providing the process transparency may be deemed unnecessary for AutoML platforms facing ordinary users. \add{To work around this limitation}, participants resort to manually tracking AutoML's learning curves, \add{which are significant indicators of its optimization trajectories}: 

\begin{quote}
    \emph{``We mainly look at how the three learning curves of training, validation, and testing change during the training process and the testing process. It may be the parameter settings or hyperparameter settings chosen for each set of AutoML. We check whether these learning curves make sense. If it makes sense, we will probably trust these results.''} (P3)
\end{quote}

\del{P3 highlighted the learning curves of model training when evaluating the results of AutoML. Learning curves are significant indicators for the evaluation of ML models. And with its help, users can get a better understanding of the AutoML model and decide whether an AutoML model's results are trustworthy.}

Another way to assess the dynamics of AutoML's process is to compare the difference among multiple runs of this process under varying settings, as P8 shared his practice of this approach: 
\begin{quote}
    \emph{``Basically, I will look at the statistical results, but I could maybe randomly sample several searches, and that's where I got the AutoML algorithms and performance metrics like accuracy or latency, and I try to measure the differences among different algorithms or different neurons. I think it is a very straightforward way to evaluate the performance of AutoML.''} (P8)
\end{quote}

\del{P8 reflected that it often requires not only examining the final performance summary but also comparing the results of multiple searches under varying settings to understand the dynamics of AutoML processes.} 

\paragraph{\underline{Workaround 3: \rev{Generating explanations manually}{Creating Customized Visualization}}}

\del{finds that data scientists need to use visualization to communicate and collaborate with others such as executives within their organization.} Many  participants (P1, P2, P3, P5, P13, P15) also recognized the importance of visualization not only for understanding AutoML's inner workings but also for communicating AutoML's outcomes to internal (e.g., team members and executives) and external (e.g., clients and stakeholders) parties: 
\begin{quote}
    \emph{``Our company's internal AutoML platform has a function that I particularly like, it can visualize its running process, especially when there are so many tasks, it can tell us the running conditions of each task, and also tell us the overall comparisons by showing us a table that lists the differences between different tasks. This is quite useful. AutoML is no longer a black box; it can give us some insight that helps us to reduce the unnecessary search space of hyperparameters for this kind of experiment.''} (P15)
    
\emph{``For my current job, we don’t have many new models, because companies like us are relatively stable. But if we have relatively new models or features, we will need to explain them to the clients.''} (P2)

\emph{``Personally, I don't use visualization very much. I just observe some specific numbers directly. However, if we need to report to clients, it's best to visualize it.''} (P3)
\end{quote}

\add{However, this functionality is often underdeveloped or even absent on many AutoML platforms.}\del{P2 emphasized the significance of visualization in communicating the results of AutoML to external parties, i.e., clients and stakeholders. Although the AutoML tools that they used did have a} Even when the existing visualization is sufficient for internal communication with experts, it often falls short to convey consumable information \add{to external, non-expert parties who lack relevant backgrounds}.\del{in a proper way}\del{In other words, despite that the visualization tool of the AutoML could bring certain transparency to the internal members, it failed to bring enough transparency to the outside.} \add{To work around this limitation}, participants often need to manually visualize AutoML's outcomes based on a set of pre-identified requirements to facilitate communication with external parties.

\begin{quote}
    \emph{``We have to visualize the explanations manually based on the results we got from AutoML, as the visualization auto-generated by AutoML is ugly, not informative, and not easy to understand $\ldots$ We need to make it easy to understand and look professional. There are certain requirements, such as avoiding text-heavy explanations and using more pictures. But internally, for example, within our group, we generally do not need manually created visualization, we can just use whatever features the platforms provide.''} (P2)
    
    \emph{``I manually visualize feature selection, which features are more important, the learning process, and learning curve changes in the performance of each model I trained, as the existing AutoML visualization function is simply not helpful.''} (P3)
\end{quote}

\del{Similar to P2, P3 believed that for internal communication, visualization was not that important compared to explaining to clients or stakeholders. For internal communication, the existing metrics provided in AutoML results were enough for the experts to understand and communicate with each other, but there was more information, such as changes in the learning curve, to visualize when explaining the AutoML results to clients, and they had to manually develop visualizations to include this information missing in the AutoML visualization tools.}

Participants also recognized that \add{one major challenge to creating such explanations is to visualize information in an essential but not overwhelming manner, especially when communicating with external parties with limited expertise:} 

\del{the results of AutoML could be very difficult to understand for people without relevant expertise or with limited expertise. Therefore, when manually creating explanations to explain the results to external parties, avoiding jargon or technical terms is critical. P5 further pointed out that explaining the right information and explaining in the right way are both important for external communication:}
\begin{quote}
    \emph{``Convincing people is not that easy from a technical perspective, but if you tell clients too many technical details like why a particular feature has a value of 0.7, people are going to be confused even more, so it's important to strike a balance to provide just enough information in the right way, not too much, not too technical, or too detailed. This can also protect us by preventing other people from copying our idea.''} (P5)
\end{quote}

\subsubsection{\bf Mitigating Potential Privacy Risks}
\label{sec:privacy}
In addition to AutoML's functionality limitations, many participants (P1, P3, P4, P5, P6, P9, P10, P11, P13, P15, P17, P18, P19) also expressed serious concerns about potential data privacy issues in using AutoML platforms. 

One major concern is whether using AutoML platforms may entail the privacy leakage of training data, which is especially consequential for critical domains (e.g., healthcare and insurance) involving sensitive information such as health history, credit history, and demographic information:

\begin{quote}
    \emph{``The problem of data privacy is quite serious. Some projects I have done before were related to medical information, which involved patient data $\ldots$ The initial data may first be provided by the hospital itself. However, if an institution provides you with a very small amount of data, while we need to do this experiment on a large scale, we must involve patient data provided by different institutions or different hospitals. Then there are privacy risks: First of all, the patient’s information cannot be leaked. Second, each hospital may not want its data to be somehow leaked to other hospitals. So privacy is definitely a very important part to consider when it comes to whether to use AutoML or which one to use.''} (P18)
\end{quote}
\del{Given the sensitive nature of the medical and healthcare fields in which he is working, P18 expressed serious concerns about the data privacy risks when using AutoML, which are the key factors for him to decide whether to use AutoML or choose which AutoML platforms to use.}

Another major concern is whether AutoML-generated ML solutions are subject to potential inference attacks if disclosed to and used by unauthorized parties. For instance, the models generated by AutoML based on confidential medical data carry a significant amount of sensitive information from the original data, while malicious parties, if given access, may infer such sensitive information by reverse-engineering the models:

\begin{quote}
    \emph{``If I get a parameter after training on the data of a bank or a hospital, I want to use it in the parameter space of other hospitals or other banks. Sometimes the parameter itself can be used to infer what the previously trained dataset looks like, which may cause data leakage.''} (P18)
\end{quote}

\del{P18 indicated an important security concern that the ML solution generated by AutoML may be subject to inference attacks, which may cause severe privacy leakage.} 

\add{To alleviate the privacy concerns above,} participants resort to various workarounds as detailed below.

\paragraph{\underline{Workaround 1: \rev{Limiting}{Uprooting} Privacy Leakage\del{at its root}}}
One straightforward workaround by participants (P6, P10, P13, P15, P17, P18, P19) is to limit privacy leakage at its root. This strategy can be adopted at either the user or organization level. Specifically, at the user level, they purposely collect less sensitive data during the data collection stage before using such data on AutoML platforms:

\begin{quote}
   \emph{``We are only collecting minimal identifiable data right now such as users' phone numbers because that's how we engage with the users over SMS (Short Message Service). We also have information about health facilities or hospitals when users sign up. [This information] is enough for us to help the users connect to health services, but we are not collecting other information such as user names, addresses, or ages.''} (P6)
\end{quote}

In addition, participants (P13, P15, P17) also mentioned that their organizations may have already performed certain data anonymization to protect data privacy before handing over the data:
\begin{quote}
    \emph{``If our company asked us to compress a model, we won't have too many images due to user privacy, or we may have a lot of data, but we do not have sensitive information such as gender since such information has been masked by the company.''} (P15)

    \emph{``Basically, the data I get may already be processed in advance. The information that can be stored in the general database is basically not related to any personal information.''} (P17)
    
     \emph{``Clients in the healthcare industry, for example, a pharmaceutical company we have collaborated with before, have strong compliance requirements, so they will also do a lot of  processing on their side.''} (P13)
\end{quote}

\del{P13 suggests that in certain areas, such as healthcare, where privacy issues are particularly important, their clients will ensure that data has been processed rigorously, alleviating their concerns about privacy breaches. 
The above-mentioned participants shared their approaches to privacy issues they have addressed using data anonymization.} 

\add{In general, only uploading non-sensitive data to AutoML platforms greatly reduces the risks of privacy breaches during AutoML's process. On the downside, this strategy may significantly affect data authenticity and negatively impact AutoML's performance, as noted by P11:}
\del{Also, P$_{11}$ pointed out another benefit of using internal platforms is that it helps preserve the data authenticity:}
\begin{quote}
    \emph{``Google definitely doesn't want users to worry that their data will be leaked, so they (Google) may mask some data and may use other means, such as protecting the user's ranking layer to protect data privacy, but such techniques actually damage the authenticity of the original data and affect performance.''} (P11)
\end{quote}

\paragraph{\underline{Workaround 2: Applying Privacy-Preserving Techniques}}
Another workaround mentioned by participants (P3, P9) is to proactively apply privacy-preserving techniques during AutoML's process. Examples include ``black-box optimization''~\cite{bbo} that avoids direct access to data, and ``federated learning''~\cite{yang2019federated} that constructs ML models using data spread across multiple parties yet without sharing data between different parties: 
\begin{quote}
    \emph{``There are some black-box optimization methods that AutoML does not touch [your data in optimization]. In this case, it can be done at least a little better to guarantee privacy. Another way is through federated learning, which is equivalent to giving data to local users without uploading [the data to the server]. It relies on the local side to do some [AutoML] searches. What AutoML receives is some high-level [data] or metadata instead of data from users' own devices.''} (P3)
\end{quote}

\add{However, this workaround is not for every AutoML user, as many may lack the necessary technical expertise to apply advanced privacy-preserving techniques.}

\del{P3 highlighted that users can apply various privacy-preserving techniques to avoid data leakage. With these techniques, users can avoid data leakage in the automation process.}

\paragraph{\underline{Workaround 3: Delegating to Legal Regulation}}
In addition, several participants (P1, P13) who use commercial AutoML platforms referred to data privacy as a legal issue that should be clearly specified in the privacy agreements:

\begin{quote}
    \emph{``I think it is a legal issue between the platform and the company. Before the company decides to use these platforms, it must clearly state the privacy issue in the confidential agreement. If the AutoML platform violates the regulations, it will be a legal issue.''} (P1)
    
    \emph{``Before we collaborate with the companies that provide the AutoML services, we must first make it clear about what data can be shared, and to what extent the data can be shared. For such issues, these (AutoML) companies actually have their own standards and their own legal team will handle such issues.''} (P13)
\end{quote}

\del{P1 stated that the AutoML platform should be regulated by privacy agreements and protect the data that users upload to it. P13 further pointed out the importance of regulation:} 

\del{P13 mentioned that it is necessary to make it clear upfront to clients about what type of data they can share. Both P1 and P13 emphasized that all parties involved in the AutoML process should be aware of and comply with regulations to protect data privacy and security.} 
Delegating to legal regulations helps AutoML users clarify their responsibilities and secure their data \add{from a legal perspective}.

\paragraph{\underline{Workaround 4: Choosing Trustworthy Platforms}}
Participants reflected contrastive views towards the trustworthiness of cloud-based AutoML platforms in terms of privacy protection. While some participants (P2, P4, P8, P9, P17) raised concerns about the privacy risks of using cloud-based AutoML platforms, other participants (P5) trusted the AutoML services of renowned companies. For example, P4 from a healthcare company explained his strategy of avoiding using cloud-based AutoML platforms when it involves private data:
\begin{quote}
    \emph{``If using the public dataset to try out the AutoML service, I'm not worried about data privacy; if I'm going to use some private datasets, I will probably not upload the data but run it locally. If I'm using cloud AutoML services, I will not choose to upload all the private data to the cloud server.''} (P4)
\end{quote}
\del{P4 indicated using AutoML services when processing public datasets. But for private datasets, P4 would develop ML models locally or only upload part of the private data to cloud servers. Users will evaluate the data that they are using, and when they feel that data security needs attention, they will be more careful in choosing AutoML's platform. They will choose to build the model locally when the existing platform does not meet their privacy requirements} P9 also echoed P4's concerns:
\begin{quote}
\emph{``The people who really have this concern won't even need to ask, they have very strict rules to prevent them from uploading any data to the cloud so this option was rolled out at first glance, so they wouldn't need to ask and this definitely a concern for many companies they don't want to disclose their data, because they have strict rules to upload data to any other servers besides their internal servers.''} (P9) 
\end{quote}

\del{P9 noted that there are strict rules to prevent them from uploading data to any other servers besides internal servers in many companies. This suggests that concerns about data privacy may prevent the application of cloud-based AutoML in certain circumstances.} 

On the contrary, other participants, \add{especially ones from startup companies (P5)}, prefer reputed cloud-based AutoML platforms (e.g., Amazon AWS) over their own in terms of data privacy protection and\del{as their own AutoML platforms due to their lack of IT infrastructures and personnel to maintain such AutoML services. For example, the participants from startup companies express more reliance on cloud-based services from those renowned platforms such as Google AutoML and AWS because they} believe 
these large companies are better positioned to protect data privacy given their plenty of infrastructure and personnel resources: 
\begin{quote}
    \emph{``If I'm hosting service on my own server, I'll be very concerned about getting attacks, because as a start-up, we cannot afford to have an onsite, fully dedicated security team. But those bigger companies have teams of experts and engineers that can take care of this.''} (P5)
\end{quote}
\del{P5 stated that for startups, using AutoML platforms provided by large companies would instead enhance the protection of data privacy.}

Apparently, this view contradicts that of participants who choose to use internal platforms for risk control, implying the complex landscape of how users choose among different AutoML platforms, which are affected by perceived privacy risks, operational costs, and platform trustworthiness.

\subsubsection{\add{Summary}}

Acknowledging AutoML's limitations, participants exercise user agency to adjust the use of AutoML in their work including \add{contextualizing input data, incorporating domain knowledge, and building internal AutoML tools}; they apply a variety of strategies to assess and evaluate AutoML's outcomes and process to increase its transparency; meanwhile, they also create new ways to combat privacy concerns, \add{ranging from data anonymization to switching between different AutoML platforms.}

We further probed participants about where they often seek help to tackle AutoML's deficiencies. Multiple complementary resources are  mentioned, including official documentation of AutoML platforms (P1, P2, P4, P5, P7, P10, P18, P19), online forums where ML practitioners gather (P1, P2, P4, P5, P12, P17, P18, P19), online searches (P2), internal training (P11), and personal networks such as friends and colleagues who are experienced in AutoML (P1). 

\subsection{Applying AutoML \add{Selectively} and Situationally}
\label{sec:selective}
As our study showed, participants are fully aware of AutoML's deficiencies and devise various strategies to work around or mitigate such limitations. Yet, they also understand that it is impractical to completely address all the limitations. Thus, \del{they are cautious about whether to apply AutoML and when to apply on a case-by-case basis.}they perform a careful cost-benefit analysis, set clear expectations, and strategically \add{decide whether and when to apply AutoML on a case-by-case basis. Although this is not a direct way to tackle the limitations per se, it is a fundamental strategy to leverage AutoML in complex situations.}

\subsubsection{Performance-Driven (Non-)Use of AutoML} 
We found that participants set clear performance expectations and decide how to use AutoML accordingly. For example, P8 explained that he evaluates the performance of AWS AutoML before making the decision of whether to use its service: 
\begin{quote}
   \emph{``I only use AutoML as a mechanism that automatically runs the new inputs through the whole training process in the background and then presents me the results. If the results pass a certain threshold, I can automatically deploy a new model. Amazon AutoML service does a very good job in training the whole model and in all implementations on a certain API (application programming interface).''} (P8)
\end{quote}
\del{P8 highlighted that his evaluation of Amazon AutoML service's performance
was overall satisfactory on the whole process and indicated that it was a good mechanism for automating the whole process of model development. Therefore, he was willing to use Amazon AutoML in proper situations.}

In certain cases, participants (P1, P3, P5, P6, P7) acknowledged that AutoML may not produce the optimal results; however, as long as the results meet their performance expectations, they are willing to adopt AutoML for its convenience:

\begin{quote}
\emph{``It doesn't necessarily mean that I have to find a perfect model when  I have a standard for performance, I will continue using it to improve something based on it, and I don't expect anything else.''} (P3)

\emph{``In our company, we only apply AutoML to get the NLP (natural language processing) for our chatbots, and we are satisfied with 80\% accuracy or so.''} (P6)
\end{quote}

\del{Similarly, P6 explained their expectations,}

\del{P3 and P6 both acknowledged that AutoML had limitations and might not be the best model. They were willing to apply AutoML in their work as the performance of AutoML was sufficient for them.  Their evaluations and expectations of AutoML's performance played a significant role in their adoption of AutoML techniques. When AutoML met their expectations, they were willing to accept AutoML for its convenience of automating the ML process despite its other limitations.} This finding corroborated previous studies~\cite{crisan2021fits}\del{finding} that data scientists are often interested in using AutoML to produce ``good enough'' results.

\subsubsection{Task-Oriented (Non-)Use of AutoML}
According to participants, they also strategically allocate different parts of their tasks to AutoML depending on the specific contexts. For example, for P2, who works in an insurance company, one daily task is to explore the consumer complaint database and classify clients based on their comments on the financial products. Her strategy is to first run the sub-task of text classification internally using her familiar techniques and then assign the other sub-tasks to the AutoML platform: 
\begin{quote}
\emph{``We need to classify customers based on their complaint records. Part of the task is to perform a classification based on the clients' complaints. We already have a fixed model to use, which we are familiar with, and have more confidence in this process than using AutoML. We use AutoML for the rest of the task.''} (P2)
\end{quote}

\del{P2 opted to utilize AutoML for certain sub-tasks, while she completed others.} In the same vein,\del{some participants (P$_2$, P$_8$) use AutoML was just a helper for them to deal with the jobs that had clear patterns.} 
other participants (P1, P4, P7, P8, P18) set \add{clear needs in mind and} only use AutoML for sub-tasks that strictly fit AutoML's intended use (e.g., architecture search, model training, and hyper-parameter tuning):  
\begin{quote}
    \emph{``We mainly use AutoML to find specific architectures and also fit parameters.''} (P4)

    \emph{``All of those AutoML tools are that automatically help me handle those issues of speech-to-text [translation] and NLP issues, so it saves me time to deliver the app that I want to design for my users in my project.''} (P7)

    \emph{``The first thing I need to do is to define the search space for AutoML before I train models. Then I also define the search algorithm. After these, \rev{models can be trained on this dataset to get a good model. If we design these things in advance, we don't need to build the whole model ourselves but}{\ldots} just use AutoML to build the models for us.''} (P17)

    \emph{``I only need AutoML to do the hyper-parameter search.''} (P18)
\end{quote} 

\del{P17 used AutoML to build a model after he had designed the whole process from the beginning. Some other participants [P4, P18] also shared that they only used AutoML under the conditions that the processes and standards had been set. Besides, when they had an overall understanding of the whole process, they may adopt AutoML to limited steps of model development. For instance,} 

\del{Both P4 and P18 had clear needs set in mind before using AutoML. P4 mainly applied AutoML for architecture and P18 mainly for hyperparameter search, instead of using it for the whole process.} 

As another concrete example, P13 works for a company in China that provides whole-package solutions to help enterprise users, especially those from traditional industries, leverage AI and ML in their production. AutoML only accounts for a small portion of the whole solution.

\begin{quote}
   \emph{``For AutoML, it can speed up the process of searching for optimal hyper-parameter settings, but it cannot replace our early stage. The understanding of the problem, the definition of the problem, and the possibility that we keep re-designing our model according to the model architecture. Actually, there are very limited now. If the model is already trained and I want to refine it, then we may be able to use AutoML to search for such an optimal parameter of a hyperparameter. For example, when we cooperate with financial companies, they have the task of automatically searching for and selling investments, which mainly rely on AutoML. In fact, it only accounts for a part of our actual delivery.''} (P13)
\end{quote}
\del{P13 mentioned there were two reasons why only to apply AutoML in hyperparameter search: the necessity of human involvement to define and understand the problems, and the need to refine the trained models instead of searching for new models.}

While the aforementioned participants apply AutoML as part of the solutions in their tasks, other participants (P10, P17) mainly use AutoML for research purposes. For instance, 
P10 uses AutoML as a baseline check for their own models, while P17 considers AutoML as a promising research topic and focuses on improving its practical impacts:
\begin{quote}
  \emph{``I use AutoML for doing experiments like comparing the AutoML models we received from Microsoft to the one we trained; we just run some experiments on it.''} (P10)
  
   \emph{``I would consider AutoML as a research topic but would not use AutoML tools at work. The current applications of AutoML are very limited, and I do not use it for my work.  It is because the application of AutoML in practice is not very successful, and there is still room for AutoML to improve, that's why it requires more scientific research in this area.''} (P17)
\end{quote}

\del{In another case, P7, who is a researcher in the human-computer interaction (HCI) field, applied Microsoft AutoML platforms’ speech-to-text service and the NLP service as part of her research project. In this project, she designed a mobile phone application to collect user voice input to promote health outcomes and used the Microsoft Azure AutoML service to help her with text segmentation. Since the goal of her project was to improve the user experience of the application, AutoML was just applied to help with processing the initial information received from users to carry out the research. For example, P7 denoted:} 

\del{P10 told us he uses AutoML only for a baseline check for their own models:} 

Further, we also found that as \add{the functionality offered by different AutoML platforms varies}, participants often decide to use a specific platform \add{that best fits their target tasks} after performing a careful pro-con analysis:

\begin{quote}
    \emph{``My question is which one is more convenient for me. For example, Azure does punctuation recognition automatically, while Google AutoML provides some punctuation, but you need to do some work manually. I would just use Azure instead of Google AutoML.''} (P7)
\end{quote}

\del{For P7, it is significant to find the proper AutoML platform for tasks. The functionality and features offered by different platforms vary. In practice, users can make the best use of AutoML tools by choosing the platform that fits their target tasks most.}

\subsubsection{Context-Specific (Non-)Use of AutoML}
Despite acknowledging its benefits, many participants (P3, P9, P14, P15, P17) also expressed concerns about using AutoML in high-stake contexts (e.g., healthcare).\del{such as health, despite the acknowledgment of AutoML's benefits.} In fact, participants who work in the healthcare industry indicated that they tend to avoid using AutoML but still rely on human-designed models and features for multiple reasons. First, as health data is often highly noisy and complicated, its processing requires domain knowledge and past experience. Second, the cost can be prohibitive if they switch from traditional methods to AutoML. For example, P17 described the current practices of the healthcare company he works for: 
\begin{quote}
    \emph{``In our company, many features are still manually designed and processed with very traditional ML models. In fact, our company does not use the deep learning method, let alone AutoML, because the traditional methods may be more stable. In addition, traditional methods have been used for a long time; they are easier for people to use and may achieve better results. The feature engineering process in health data can be very complicated, and many experts who are already very experienced in this field can better identify this type of data.''} (P17)
\end{quote}
\del{P17 highlighted that compared to AutoML, traditional methods may be more stable. Besides, due to the complexity of health data, it is vital to incorporate experts' domain knowledge into the design of ML models. AutoML may have bad performance on some datasets or in high-stake contexts. In these cases, users may choose not to use AutoML due to practical considerations.}

Furthermore, these non-use cases are not derived from AutoML's performance issues only, but rather reflect an even larger topic of trust issues in AI or automated systems in general, known as ``algorithm aversion''~\cite{dietvorst2015algorithmaversion}. For example, both P15 and P3 expressed concerns about the reliability and trustworthiness of AutoML in critical domains (e.g., medical analysis) and emphasized the importance of human expertise: 
\begin{quote}
\emph{``People can still not be replaced by machines in some very critical scenarios, such as doing some analysis of medical treatment. For example, read a CT scan or some X-rays to determine whether this person has lung cancer. I think such a thing is difficult to replace by AutoML because it requires human knowledge and the cost will be too high if AI makes a mistake.''} (P15)

\emph{``In some medical situations, in which some data itself is more sensitive, many people do not trust ML models designed by people, let alone the ones designed by AutoML. For example, you have 99\% human inspections. If it is true that there is a problem with this model in 1\% of practical applications, you have no way to hold it accountable.''} (P3)
\end{quote}

\del{P15 expressed concerns about the reliability of AutoML in critical scenarios such as medical analysis. Users believe that in some domains automated models cannot replace human expertise. Therefore, in some high-risk task scenarios, users may avoid using AutoML. P15's view was echoed by other participants. For example, P3 added:}

\del{P3 expressed the distrust in AI in general and the ``algorithm aversion'' in the health context. This reflects the fact that users' attitudes toward AI in general and their domains can greatly affect their attitudes and level of use of AutoML.} 

We further probed the reasons behind the distrust in AutoML and found that it is partially due to the challenge of holding AutoML accountable. According to participants (P9, P14), when AutoML is involved in the decision-making process, it is difficult to attribute responsibility when an error occurs or when AutoML's performance falls short of expectations. As P14 explained: 
\begin{quote}
     \emph{``I think things that are simple and highly repetitive can be handed over to AutoML, but the more critical parts, such as decision-making or analysis, still need people because these tools cannot take responsibility. A very key problem is to what extent the achievements of these AI tools represent humans' achievements. This needs to be carefully defined because what AI does is not necessarily what humans will do.''} (P14)
\end{quote}
\del{P14 emphasized that AI tools cannot fully take humans' responsibility and represent humans' achievements. This indicates that users have concerns about the attribution of responsibility when adopting AutoML and thus may choose not to involve it in the critical parts of modeling.}

\subsubsection{\add{Summary}}
\add{As our study showed, besides working around AutoML's inadequacies, participants also strategically decide whether and when to use AutoML based on their} performance-driven, task-oriented, and context-oriented motivations, and pragmatically adjust their use of AutoML to fit their needs and background knowledge.

\section{Discussion}
Our study reveals the highly heterogeneous nature of how users incorporate AutoML as available resources to help them accomplish their tasks and goals in practice. Users often develop different understandings and expectations of AutoML's capabilities, strategically adjust their use of AutoML technologies and develop pragmatic workarounds when facing challenges.\del{brought by AutoML's lack of customization, lack of transparency, and privacy issues.} Previous work has mentioned similar challenges (e.g., customizability~\cite{xin2021whither} and transparency~\cite{wang2019atmseer,crisan2021fits}) and suggested that resolving these issues requires human intervention. Our study extends previous work by elucidating users' efforts to overcome specific challenges associated with AutoML and balance AutoML with other resources in practice. In addition, our study discovers privacy concerns as a novel issue associated with AutoML use, which has yet to be discussed in the current discourse of AutoML in HCI. Our study unpacks users, as situated actors, who develop workarounds in overcoming the challenges and examines the underlying decision-making process. 

\subsection{\add{User Agency and Workarounds}}
\add{According to the workarounds theory~\cite{alter2014workarounds}, ``workarounds are fundamentally about human agency, the ability of people to make choices related to acting in the world.'' In other words, human agency manifests in the development and execution of workarounds to meet individual needs, goals, and expectations~\cite{kaptelinin2006acting} and users often define their goals according to complex situations and use different resources to accomplish them~\cite{suchman1987plans}. In our study, users' workarounds reflect their needs to take control over AutoML and to restore user agency when facing constraints, challenges, and unmet expectations derived from technological and situational factors. Below, we detail the active role of user agency in this decision-making process, specifically, how to develop workarounds and how to decide (non-)use of AutoML.}
\del{the process of exercising user agency in overcoming challenges brought by AutoML, which involve situated actions based on cognitive evaluations, such as balancing available resources, expertise level, and specific goals. We also found that the extent to which users seek control and trust AutoML is affected by attitudinal factors such as belief in automated systems.}

\subsubsection{\add{User Agency in Developing Workarounds}}
\add{The activity theory~\cite{wertsch1998mind} proposes that users ``appropriate'' tools to empower them to achieve goals and they often need to combine multiple tools to do so~\cite{kaptelinin2006acting}. In our context, how users exercise their agency is reflected in their developing, selecting, and executing workarounds. Further, we observed that the level of user agency hinges upon external factors such as resources available as well as internal factors such as users' ML expertise.} For example, while ``power users'', or users with extensive ML experience, often apply more advanced techniques to verify AutoML's performance and leverage various tools to customize the use of AutoML, users with less ML expertise rely more on the available features of AutoML platforms and are less likely to enact workarounds. \add{This finding demonstrates how users ``use their available knowledge to create and execute an alternate path to achieve the goal'' ~\cite{alter2014workarounds}}, and corroborates previous HCI research on that experienced users are superior at recognizing design defaults~\cite{dreyfus1986power}, comprehending functional relationships~\cite{chi1981categorization}, and developing problem-solving strategies~\cite{klein2017sources}. Therefore, for users with less ML knowledge, we expect an even narrower range of enactment as AutoML solutions become more tightly integrated into the data science workflow. As previous studies reflected that lay users tend to overly rely on ML~\cite{chiang2022exploring,chiang2021you},\del{where ML literacy plays a critical role here~\cite{chiang2022exploring},} we should be cautious.\del{as users become more dependent on using integrated technologies such as AutoML.} Although AutoML is seen as a solution to accelerate ``democratizing data science''~\cite{shang2019democratizing}, \del{applying it correctly and effectively requires a certain level of knowledge and expertise in ML.}
limited knowledge, experience, or expertise in ML could result in over-trust and over-reliance\del{on the automated services of AutoML}, which may threaten user agency as users develop cognitive heuristics or mental shortcuts to perceive and use AutoML~\cite{sundar2020rise}.\del{Especially With the current limitations of AutoML, it is critical to consider ways to foster learning and develop effective workarounds.} 

\subsubsection{\add{User Agency in (Non-)Use of AutoML}} 
\add{Based on the activity theory~\cite{kaptelinin2006acting}, users' goals, interests, and intentions are the starting points for analyzing situations in problem-solving.} In our study, participants perceive AutoML services as part of the tools to accomplish the tasks at hand~\cite{karmaker2021automl}. Further, according to the expectation confirmation theory~\cite{bhattacherjee2001ECT}, the continuous use of technology is largely determined by perceived expectation confirmation. Our findings show that participants' use of AutoML depends on their evaluations of AutoML's capabilities and performance expectations. As long as AutoML's performance and functionality meet such expectations, they will use AutoML in practice. This view is different from previous studies that examined the reactive role of humans from the technical perspective of data science workflows~\cite{xin2021whither,crisan2021fits}. From our perspective, users play a more proactive\del{instead of reactive} role when considering the relationship between AutoML and data science tasks.  

In addition, some participants mentioned that they avoid using AutoML in high-stake contexts such as healthcare while having the tendency to trust humans over AutoML and ML in general. \add{In HCI, this phenomenon of ``non-use'' is also a matter of users' own choice that resemblances the sense of agency~\cite{skeba2021haschoice}.} Empirical research \add{offers possible explanations for such decisions. One explanation is ``algorithm aversion''}\del{has found that some users may hold the tendency of ``algorithm aversion''} that describes users' predisposition that favors humans over automated systems~\cite{dietvorst2015algorithmaversion}. Another possible explanation is blame attribution, which is widely discussed in human-AI collaboration, especially when moral responsibility is involved~\cite{arntz2021influence}. \add{Therefore, such pre-existing attitudes toward automation, in general, could drive the decision-making towards the non-use of AutoML.}

\subsection{Designing to Mitigate AutoML's Limitations}
\del{Xanthopoulous et al.~\cite{xanthopoulos2020putting} proposed a user-centric framework for comparing AutoML services, including six categories: Estimates (``the wealth and depth of estimated quantities regarding the predictive model''), Scope (``the range of input data that can be analyzed''), Productivity (``the ease of use and boost of user productivity''), Interpretability (``data visualization'', ``progress report'' and ``final model interpretation''), Customizability (``the ability of the services to customize analysis according to user choices and preferences''), and Connectivity (``connecting with external tools and resources''). Existing HCI research on AutoML research in HCI has largely fallen into these metrics which primarily focus on AutoML's performance and functionality.} Our study corroborated existing HCI research on AutoML  (e.g., \cite{xin2021whither, crisan2021fits, wang2019atmseer}) by confirming that the lack of customizability and transparency are users' two major concerns. In addition, our study also uncovered privacy as an additional significant, non-functional concern, which has not been identified in previous research. Moreover, our study detailed the workaround strategies and users' rationale and practices of applying AutoML selectively and situationally. Below we discuss what insights our findings yield in terms of designing to mitigate the three perceived limitations of AutoML.

\subsubsection{Supporting Domain-Specific Customizability}
AutoML's lack of customizability is one major limitation perceived by participants. Previous studies reported a similar issue but emphasized the need for more user control over AutoML's process~\cite{xin2021whither, crisan2021fits}.\del{Thus, increasing attention has been devoted to enriching customizable options for supporting AutoML tasks individual segments of the data science workflow (e.g., optimization metrics and search space)~\cite{wang2021flaml}.} However, this effort to improve AutoML customizability focuses mainly on giving the user control over each stage of the standard data science workflow due to the tension between full automation and human intervention as identified by the ``human-in-the-loop'' approach~\cite{wang2021flaml}. Our study pointed out that, in addition to providing workflow-focused customizability, it is also important to design for domain-specific customizability.

Our study found that the lack of customizability also derives from the tension between AutoML as the main service and the support for inclusive services for users with special domain needs. Participants frequently referred to AutoML's lack of customizability as its incapability to fit domain-specific situations. This finding calls for more attention to AutoML platforms that are specialized for concrete application domains and can be adapted to specific contexts~\cite{haghighatlari2020chemml}.\del{One example is that, in the field of computational molecular science, researchers have developed an open ML and informatics program suite that supports data-driven research in the chemical and materials domains with customizable features~\cite{haghighatlari2020chemml}. More development effort like this is needed.} HCI research may consider investigating how to design such customization in a user-friendly manner.

\subsubsection{Providing Multifaceted Transparency}
The second challenge we found for AutoML is its lack of transparency. Previous research has also pointed out this issue, but focused mainly on user evaluation of current visualization tools~\cite{crisan2021fits} or the development of new visualization techniques to improve user understanding of AutoML's outcomes~\cite{drozdal2020trust, wang2019atmseer, weidele2020autoaiviz, narkar2021model}. Our results reflected a more nuanced need for AutoML's transparency: users tend to favor different transparency features in evaluating the outcomes and/or process of AutoML. Some users rely on the performance metrics provided by AutoML platforms, while others need to comprehend AutoML's training and selection processes.\del{In fact, how to employ different statistical metrics to evaluate classification in practice is still under debate~\cite{chicco2020advantages}. Users may need different metrics according to different tasks, such as F1 scores when having unbalanced data. Additionally, consumers demonstrate a lack of understanding of the models themselves.} Such findings corroborated explainable AI (XAI) research indicating that there is no one-size-fits-all solution for transparency and that a more personalized, interactive method of interpreting ML to users that supports user requirements is often necessary~\cite{earp2016personalized,liao2020questioning, sun2022exploring}.

Additionally, current AutoML transparency tools are designed only to support internal validation and understanding of AutoML solutions, while user expectations go beyond this scope. Our study revealed that AutoML users require transparency features not just for internal evaluation (e.g., model explanations) but also for external communication with clients, which raises the requirement for layman's comprehension through easy-to-understand visualization. This finding echoed previous work that showed that data scientists tend to have high expectations of what transparency tools can achieve that are beyond the tools' capabilities~\cite{kaur2020interpreting}. The gap between user expectations of what AI can do and the designed features in reality may undermine user trust and adoption of AI~\cite{kocielnik2019will}. To enhance transparency, it is therefore imperative to consider different stakeholders involved in the process of understanding AutoML~\cite{preece2018stakeholders}.

\subsubsection{Enhancing Data Privacy}
Moreover, a new issue of privacy surrounding AutoML has surfaced, which has not yet been discussed in the current HCI literature on AutoML. The striding advances in ML techniques, especially deep learning techniques, are built upon the consumption of massive amounts of data, which are often sensitive by nature, leading to unprecedented privacy concerns. Prior work has investigated possible privacy challenges and risks arising in the ML process~\cite{ml-memory,model-stealing,membership-inference}, showing that both ML models and training data can be the targets of privacy attacks and leak sensitive information. For instance, Fredrikson et al.~\cite{fredrikson2015model} demonstrated that even if ML service providers (e.g., Google) provide prediction capabilities as query-only services, the attacker is able to reconstruct the confidential ML models via querying such services in a black-box manner.\del{Compared to the plethora of studies on the privacy of ML processes, privacy issues in the context of AutoML are much less understood.} Besides the privacy issues in common with general ML (e.g., the requirement for massive amounts of training data), AutoML also incurs unique privacy risks and challenges. For instance, Pang et al.~\cite{pang2022security} showed that AutoML-generated models tend to more easily leak sensitive information about training data. Additionally, the practice of AutoML brings in new stakeholders (e.g., AutoML service providers), entailing privacy risks unaccounted for in the literature.\del{In this study, we aim to understand, from the perspective of AutoML end users, how they perceive and mitigate such privacy risks.} 

Our study found that most participants concern about privacy while their coping strategies range from non-use of AutoML platforms to privacy-preserving techniques. They indicated that they use AutoML with privacy in mind but do not see it as a reason to avoid using AutoML as they feel that AutoML's perceived benefits \add{(e.g., dependable and fast services)} outweigh its potential privacy risks.\del{Some participants, for instance, voiced worries over data privacy, but the benefits of using huge platforms such as Google AutoML and AWS exceed the potential risks since they offer more dependable and quick service.} Laufer et al.~\cite{laufer1977privacy} referred to this behavior as privacy calculus, a cognitive process in which people estimate future outcomes of current decisions by calculating the costs and advantages of sacrificing some privacy for better outcomes. Previous HCI research has focused on technological or behavioral mechanisms as privacy coping techniques, which are defined as active problem solving based on deliberate cognitive assessments~\cite{laufer1977privacy}. These coping strategies include evaluating privacy policies~\cite{karat2006evaluating}, adopting privacy-enhancing tools~\cite{yu2016iprivacy}, and information withholding~\cite{gambino2016user}.

\subsection{Supporting Collaborative Work behind AutoML}
Previous work has shown that substantial human work is necessary to apply AutoML in enterprise contexts~\cite{crisan2021fits}. Humans are ``valuable contributors, mentors, and supervisors to AutoML'' by providing guidance throughout the data science workflow~\cite{xin2021whither}. Our study found that the need for human support goes beyond this scope, necessitating collaborations among ML experts, data scientists, and domain experts for the successful adoption of AutoML. 

\del{Previous research on ML has emphasized that a common issue with ML tools is that potential users who are often domain experts have limited involvement in the development of ML models \cite{amershi2014power}. To resolve this issue, Gil et al.~\cite{gil2019towards} have proposed the ``human-guided ML'' approach that designs systems to support direct input from domain experts. However, the} Existing AutoML platforms are often designed for generic ML problems and lack capabilities to integrate domain knowledge. To bridge this gap, data scientists could play the role of helping translate domain knowledge into AutoML's process. Existing research has shown that data scientists increasingly work with domain experts to solve complex scientific problems ~\cite{mao2019data}. Our study also corroborated this finding, as stated by participants (P13), that the ``bridging role'' of data scientists is critical for helping translate domain knowledge into AutoML. 

In addition, our findings surfaced how teams within organizations collaborate\del{--oftentimes, it is not a single-user scenario}. For example, to remedy AutoML's privacy issues, it often requires the company's legal team to assess the privacy compliance of AutoML services; also, technical teams need to develop intuitive visualization to explain the ML solutions generated by AutoML to other stakeholders.\del{Such findings are consistent with previous literature that reveals that visualization tools are often needed to support collaborative data work between different teams \cite{crisan2021fits, passi2018trust}.} In theory, a human-in-the-loop paradigm for augmenting the data science workflow can be useful for understanding the types of engagement between humans and machines to ameliorate certain trust concerns. However, we found that human-in-the-loop is limiting since an AutoML correction and refinement loop not only exists within a wider scope of data science processes but also within organizational processes. While the nomenclature of human-in-the-loop is not exclusive to single individuals interacting with AutoML, we argue that the notion of ``{\em humans}-in-the-loops'' more accurately captures how AutoML technologies are used within enterprise settings~\cite{crisan2021fits}.   

\subsection{Design Implications}

\del{The findings of this study seek to present a holistic view of how users exercise agency when working with and around AutoML technologies.} First, our study highlighted the heterogeneous nature of real-world data science. Participants reflected that current AutoML platforms are not customizable for certain domains (P7) or have not addressed language localization (P6). One practical implication is to design domain-specific AutoML platforms (e.g., healthcare~\cite{waring2020automated} and finance~\cite{agrapetidou2021automl}). Another way to improve compatibility with specific contexts is to enable users to select relevant tasks or contexts~\cite{karmaker2021automl}. Techniques, such as active learning, that help AutoML capture user preferences may be promising. \add{Also, techniques such as backward compatible learning~\cite{hu2022learning, shen2020towards} may help ensure the backward compatibility of AutoML-generated solutions.} As participants indicated, domain knowledge is also vital for resolving the\del{AutoML non-} customizability issue; thus, it is crucial to develop features that incorporate domain knowledge to support context-specific applications.\del{and practical applicability in a business context.} For example, it may allow users to specify domain knowledge as first-order relations or introduce connections into neural network models based on the logical constraints enforced by such domain relations~\cite{first-order-logic}; it may also allow users to refer to domain-specific sources (e.g., knowledge graphs) to perform data augmentation before feeding the data to AutoML~\cite{krisp}. 

Meanwhile,\del{considering the varying focus on what should be interpreted and to whom it should be conveyed,} our research discovered that the present transparency tools do not yet fulfill the needs of AutoML users. Although some participants are satisfied with the transparency of AutoML's outcomes, others address more concerns about understanding and monitoring AutoML's process. The design of future transparency tools could consider adding different features to provide both static (i.e., results) and dynamic (i.e., process) transparency. In addition, as some participants reported, there is an increasing need to communicate AutoML results to other stakeholders, such as clients without relevant ML backgrounds, transparency designs could include not only unpacking algorithmic black-boxes, but also addressing how to communicate data and models among teams, products, and services. 

Lastly, our research also highlighted the necessity of resolving privacy concerns when implementing AutoML in real-world contexts, which echoed the call for addressing privacy by design~\cite{wong2019bringing}. Specifically, our study found that participants with more extensive ML expertise tend to show a higher level of privacy awareness. On the other hand, participants with less expertise and those from enterprises with fewer resources rely on the legal compliance of AutoML platforms. Thus, designers of AutoML platforms could consider more proactive approaches, such as incorporating privacy notices~\cite{egelman2009timing} and nudges~\cite{chang2016engineering} to make users aware of privacy risks and engage in privacy-enhancing practices.

\section{Limitations and Future Work}
The study has a few limitations that could inform future research. We discovered that our participants come from a wide variety of backgrounds and have varying understandings of AutoML. Despite the fact that our findings shed fresh light on how users actively adopt AutoML in real-world contexts, our participants are not sufficiently representative to permit comparisons. For example, \add{we recruited our participants mainly through words of mouth, mailing lists within enterprises, and social media, with diversity potentially limited by the nature of such channels.} Future research could examine the difference between AutoML users with more varying ML expertise levels or from more diverse domains to further elucidate how their competence and domain needs may impact their perceived challenges and workarounds. \add{Future studies could also expand on the results of our study through quantitative research methods that lead to more generalizable implications.} \add{Furthermore, the voluntary participation in our study may result in self-selection bias ~\cite{robinson2014sampling}, with participants who consented to participate in our interviews may be more active users compared to those who did not. This opens the questions for future research on comparing active users with non-active users of AutoML.} In addition, although our study discovered three main challenges that users develop strategies to cope with (i.e., customizability, transparency, and privacy), future work is needed to  systematically investigate other emerging concerns and challenges not covered in this study. For example, how AutoML may magnify the fairness and bias concerns in ML \cite{mehrabi2021bias} and how users perceive and seek solutions for such difficulties are potential topics that 
could be explored in future research. Lastly, participants outside the United States revealed novel issues such as lacking local language support from AutoML. \add{Future research could compare AutoML use cases across different countries and examine other factors that may impact the adoption of AutoML (e.g., languages and cultures).} 

\section{Conclusion}
In this research, we demonstrated how privacy concerns, lack of transparency, and lack of customizability in AutoML affect and complicate real-world data science activities. The study revealed a range of tactics used by AutoML users to control, resolve, and utilize various resources (such as internal AutoML resources, legal teams, and manual checking) to overcome those obstacles in imperfect but ultimately practical workarounds. Understanding the situated and discretionary adoption and use of AutoML opens up new possibilities for research and practices to promote more effective human-automation collaboration in applied data science. 

\begin{acks}
We thank our participants for sharing their thoughts and experiences. We also thank the anonymous reviewers for their valuable feedback. This work is supported by the National Science Foundation under Grant No. 2212323, 1951729, and 1953893.
\end{acks}



\bibliographystyle{ACM-Reference-Format}
\bibliography{main}


\end{document}